\documentclass{sig-alternate}
\usepackage{mathptmx} 
\usepackage{multirow}
\usepackage{fancyhdr}
\usepackage[normalem]{ulem}
\usepackage[hyphens]{url}
\usepackage[sort,nocompress]{cite}
\usepackage[final]{microtype}
\usepackage[keeplastbox]{flushend}
\usepackage[bookmarks=true,breaklinks=true,letterpaper=true,colorlinks,citecolor=blue,linkcolor=blue,urlcolor=blue]{hyperref}

\emergencystretch=\maxdimen
\usepackage{xcolor}
\usepackage{amsmath}
\usepackage{amsfonts}
\usepackage{authblk}

\title{FAT-PIM: Fault-Tolerant Processing-In-Memory} 
\author[1]{Kazi Abu Zubair}
\author[2]{Sumit Kumar Jha}
\author[3]{David Mohaisen}
\author[4]{Clayton Hughes}
\author[5]{Amro Awad}
\affil[1,5]{Electrical and Computer Engineering, North Carolina State University}
\affil[2]{Computer Science, University of Texas at San Antonio}
\affil[3]{Computer Science, University of Central Florida}
\affil[4]{Scalable Computer Architectures, Sandia National Laboratories}

\begin{document}
\maketitle
\pagestyle{plain}


\begin{abstract}

Processing In Memory (PIM) accelerators are promising because they can provide massive parallelization and high efficiency in multiple application domains. These architectures can produce near-instantaneous results over wide data streams, allowing for real-time performance in data-intensive workloads. For instance, Resistive Memory (ReRAM) based PIM architectures are widely known for their inherent dot-product computation capability. While the performance of these architectures is appealing, reliability and accuracy are also important, especially in mission-critical real-time systems. Unfortunately, PIM architectures have a fundamental limitation in guaranteeing error-free operation. As a result, the current methods must pay high implementation costs or performance penalties to achieve reliable execution in the PIM accelerator. In this paper, we make a fundamental observation of this reliability limitation of ReRAM based PIM architecture. Accordingly, we propose a novel solution--\textbf{Fa}ult \textbf{T}olerant \textbf{PIM} or \textbf{FAT-PIM}, that allows for low-cost error detection. Our evaluation using simulation technique shows that we can detect all errors with only 4.9\% performance cost and 3.9\% storage overhead. 

\end{abstract}

\section{Introduction}

Many modern applications operate over large datasets. The accelerated growth of the memory footprint of applications on one hand and the limited bandwidth between the memory and the processor on the other hand motivate the need for novel and unconventional processing architectures such as In-Memory or Near-Memory computing\cite{inmemory1,inmemory2,nearmemory1,nearmemory2}. In such a computing architecture, data can be processed directly inside the memory, minimizing the data movement between the CPU and the memory. Machine learning applications\cite{ISSAC,PipeLayer}, databases\cite{INM_database1,INM_database2}, personalised recommendation systems\cite{INM_personalized_recom1,INM_personalized_recom2}, and genomics\cite{INM_Genome1} benefit from the massive parallelization of in-memory computing.

Prior works \cite{ambit,seshadri2019dram,mutlu2019processing,zabihi2018memory,angizi2019mrima,li2016pinatubo} have shown that a simple change in the memory circuitry can enable tremendous in-memory computing potential. These works have proposed many ideas to allow in-memory computation in various memory technologies such as SRAM\cite{yin2020xnor,lee2020bit}, DRAM\cite{ambit,mutlu2019processing}, ReRAM\cite{PipeLayer,ISSAC}, and STT-MRAM\cite{angizi2019mrima}. For instance, activating two wordlines in DRAM allows charge sharing among capacitors. The direction of the current in the bitlines during this period can help determine the result of bit-wise operations among values stored in two word lines \cite{ambit}. Among such Processing In Memory (PIM) architectures, Resistive RAM (ReRAM) crossbars have also shown significant in-memory computation potential. In particular, they are known for their inherent capability to process matrix-vector multiplication and dot-product operations. For instance, a multiplication of a matrix and a vector can be obtained in the bit lines in the form of analog value of currents by programming the crossbar cells as per the matrix and applying the input vector through the word lines\cite{ISSAC}. With such PIM accelerators, thousands of matrix multiplications are processed in a single memory read cycle. This will achieve tremendous benefits over the traditional Von-Neumann architectures where frequent data movement between the memory and the CPU core is necessary.


Besides performance, reliability is also an important factor in PIM architectures with significant challenges. The first challenge is ensuring correctness of the input data. Due to the high degree of parallelization and data processing inside the memory, it is challenging to ensure the correctness of the massive amount of data promptly. Traditionally, the CPU reads data from memory, processes it inside its core, and finally, the computed data are pushed towards the memory hierarchy. Since there are limited degrees of parallelization, the overhead of Error Correction Code (ECC) is tolerable and does not significantly impact the performance. However, in PIM architecture, it is impractical to use a similar method and pass all data through the memory controller before they are processed in memory, as that will nullify all benefits of in-memory processing.

Nevertheless, inputs need to be checked before computation inside the memory. One possible way to achieve that is to have a scrubbing mechanism that scrubs all data periodically. Note that modern memory systems also employ ECC circuitry inside the memory DIMM for higher reliability. Similar circuitry can be utilized for data scrubbing. Such ECC hardware inside the memory is generally placed at a relatively higher level to reduce circuit overhead and design complexity (e.g., at chip-level protecting each 64b data\cite{XED}). While scrubbing can keep error accumulation in check, soft errors can occur any time and produce many incorrect results before the next scrubbing phase. Due to the periodicity of scrubbing, it's likely that these incorrect results will propagate, further corrupting the computation. Therefore, unless we squash all computed values upon error detection during the scrubbing phase and preserve the capability to roll back to a previous state, the computation cannot be trusted. As the reader may guess, this will require a complex design and will have significant slowdown in the presence of memory errors. Accordingly, this approach to scrubbing will be impractical.

Alternatively, if the ECC circuitry at a higher level is shared among many processing engines (PEs) and used before every operation, the hardware complexity and cost can be reduced. However, this will incur significant stagnation if a high-degree of parallel computation is expected. Accordingly, it is impossible to quickly evaluate the data for errors before the operation unless we adopt the costly and na\"ive way of employing ECC at the lowest level (e.g., memory crossbar, or processing engines). Adding ECC circuitry at such a low level and forcing an ECC check before every operation will incur significant hardware and power costs. 

The second challenge is to ensure the correctness of the in-memory computation itself. In most in-memory computing designs, the cellular-level analog properties (e.g., current, voltage and resistance) are utilized to perform certain operations. For instance, Ambit\cite{ambit} utilizes the level of charge stored in the capacitor, Pinatubo\cite{li2016pinatubo} utilizes the resistance level, and ISSAC\cite{ISSAC} uses the analog current to determine the result of the operation. Unlike CMOS logic, such architectures are less reliable and can be influenced by external noise. While the first challenge of ensuring the input's reliability is at least addressable at the expense of high hardware cost or a compromised performance, the second challenge is more difficult to solve. As the data are processed inside the memory, the results will not have any ECC bits associated with them. Even if we assume that all inputs have some ECC that can be used to check for errors before being used as an operand, the ECC for the computed results will be unknown. In other words, if Data {\tt A} and {\tt B} have {\tt ECC(A)} and {\tt ECC(B)}, the result {\tt C = A\ \textit{op}\ B} will have no ECC unless we can calculate a homomorphic ECC, i.e., { \tt ECC(C) = ECC(A) \textit{op} ECC(B)}. In many cases, these results are fed back into the PEs for further processing for the next stages of computation. If we allow recalculation of ECC over computed data, it may spread errors and the newly calculated ECC may certify the faulty data as correct.

Despite these challenges, it is crucial to guarantee a reliable in-memory operation. Such architectures can be used to implement systems with real-time machine learning and data analysis requirement. For instance, autonomous vehicular systems need to make prompt decisions based on the data collected from their surroundings. If highly parallel in-memory compute engines are used to accelerate the computation in this domain, an error may have devastating consequences. For example, Guanpeng et al.\cite{NVIDIA} showed that a consequence of a soft error in a DNN system might lead to classifying an oncoming truck as a bird, in which case the autonomous vehicle may decide not to initiate the braking operations.

In this work, we show that it is possible to address both challenges of implementing reliable operations in ReRAM crossbar-based in-memory architectures with minor modifications inside the in-memory compute architecture. In summary, we make the following fundamental contributions:

\begin{itemize}
    \item We present the fundamental challenges in ensuring reliable PIM operation.
    
    \item We propose a simple but effective solution of using summation as a means to having homomorphic ECC operation in ReRAM crossbar-based PIM architecture.
    
\end{itemize}

We evaluate our scheme using an in-house simulator and implement our error tolerance mechanism. We also perform fault injection analysis in our simulation to analyze the reliability aspects. On average, \textbf{FAT-PIM} can offer low-cost error detection with only 4.9\% additional performance overhead and 3.9\% storage overhead. 

The rest of the paper is organized as follows. Section \ref{sec:background} discusses the related topics and backgrounds of our work. Section \ref{sec:motivation} illustrates the limitations of the existing approaches and motivates the need for having a more reliable PIM architecture. Section \ref{sec:design} discusses the design of our scheme. Section \ref{sec:methodology} shows our evaluation settings and section \ref{sec:result} presents our results. Finally, Sections \ref{sec:discussion} and \ref{sec:related} present related discussions and relevant works. 


\section{Background}
\label{sec:background}
This section discusses the topics related to In-Memory Computing, ReRAM crossbar architecture, and ReRAM-based PIM accelerators.

\subsection{Resistive RAM (ReRAM)}
\begin{figure}[h]
    \centering
    	\includegraphics[width=\columnwidth]{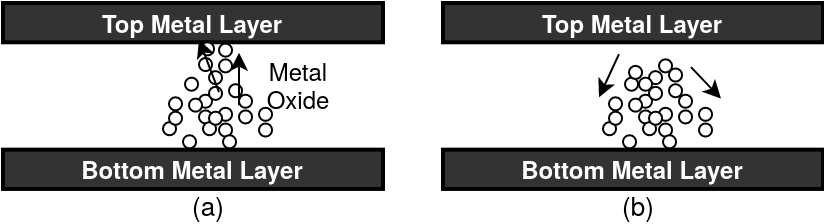}
    	\caption{ReRAM Cell (a) SET operation (b) RESET operation}
		
	 \label{fig:reramcell}
\end{figure}

Resistive memories store data in the form of resistance of the memory cell. Phase Change Memory (PCM) and ReRAM are examples of resistive memories. ReRAM usually refers to metal-oxide-based memory technology, where the oxide layer is sandwiched between two metal layers. Figure \ref{fig:reramcell} shows a simplified view of a metal-oxide ReRAM cell. The oxide layer forms a conductive filament that can be formed or raptured based on the applied voltage across two metals. The cell can retain a high-resistance state when the oxide layer is ruptured (RESET) and a low-resistance state when it forms a connection between two layers (SET). It is also possible to manipulate the density of the oxide layer by applying a varying voltage and hence vary the resistance. This resistance continuum can be used for multi-level cells.

\begin{figure*}[h]
    \centering
	    \includegraphics[width=\textwidth]{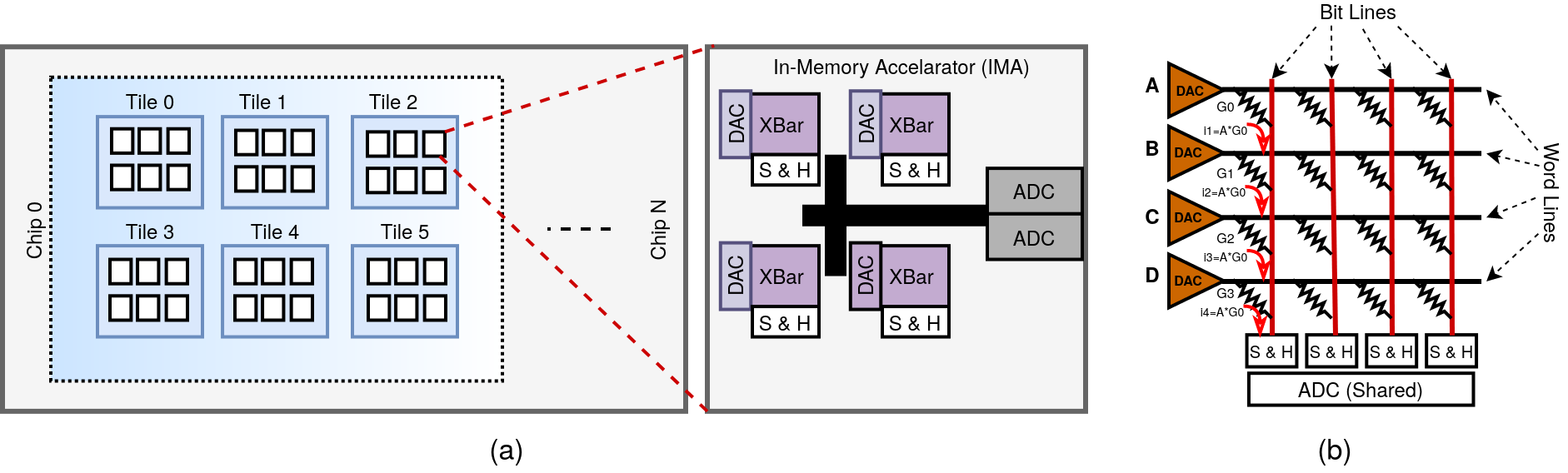}\vspace{-2mm}
		\caption{ (a) ReRAM crossbar based machine learning accelerator architecture\cite{ISSAC,PipeLayer}. (b) Dot product computation in ReRAM crossbar. }
		\vspace{-2mm}
    	 \label{fig:issac}
\end{figure*}

ReRAM can be organized in a 1T-1R fashion where a transistor controls the current through each cell. Having an access transistor in this fashion can be beneficial in terms of energy efficiency and performance\cite{cong_xbar}. However, having a transistor per cell increases the overall cell size and leads to higher area and cost. A more area-efficient way of organizing ReRAM cells is the crossbar architecture, where all cells are connected in a crossbar without an access transistor. While such structures can provide highly dense memory with minimal area overhead, they are less reliable due to the interference one cell imposes on every other cell. Since there is no access control mechanism, sneak-path currents can flow through the crossbar, causing difficulty in maintaining uniform voltage across the crossbar. When we activate a bit-line and a word-line, a small current will flow through other cells in that row and column. However, prior literature has addressed this problem by using Double-Sided Ground Biasing (DSGB)\cite{cong_xbar}.



\subsection{Memristor Crossbar-Based PIM architecture}

One exciting aspect of the memristor crossbar is that it has an inherent capability for dot product calculation, as discovered in many prior arts\cite{ISSAC,PipeLayer}. Figure \ref{fig:issac} shows such an architecture, which can produce a multiplication result between a 2D matrix and a vector in a single read cycle. Such memristor crossbar-based PIM engines primarily consists of a crossbar array, a Digital-to-Analog Converter (DAC), a Shift and Hold (S$\&$H) circuit, and an Analog-to-Digital Converter (ADC). Before the computation, the matrix (e.g., weights of a neural network) is programmed into the crossbar array in the form of conductance ($G=\frac{1}{R}$). Then, the input vector is applied through the word lines (WL) in the form of voltage. Typically, the voltages are supplied digitally, and the DAC converts them into analog voltages through the word lines. As the voltages are applied through the word lines, current starts flowing through the cells, which finally appears at the end of the bit lines in the form of a summation of currents flowing through all the cells on that bit line (BL). This sum of currents represents a vector multiplication between the applied voltage and the conductance of the cells in the bit line. Accordingly, the multiplication result between the matrix and the vector appears through the bit lines. Figure \ref{fig:issac}b illustrates this operation in simple terms. For instance, each cell will add {$ \tt A \times Gi$} current to the bit line, where {\tt A} is the input to the word line and {\tt Gi } is the conductance of the cell. In this simplified diagram, the total current flowing through the bit line will be {$ \tt I = A \times G0 + B \times G1 + C \times G2 + D \times G3 $}. 


These crossbar architectures can accelerate applications that need to perform many matrix-vector multiplication operations. For instance, in a Convolutional Neural Network (CNN), which is widely used in image recognition, computer vision, and data mining applications, convolution is the most important and repeated stage. In the convolution stage, a number of filters are applied to the input to extract specific features. This operation requires multiplying segments of the input with the filters. The ReRAM crossbar architecture can be used to perform an accelerated computation. ISAAC\cite{ISSAC} is a PIM accelerator that utilizes such a ReRAM crossbar to accelerate CNN applications. Although we aim not to specialize our design to any specific use case, FAT-PIM adopts ISAAC's underlying hardware for PIM operation and provides additional reliability. Therefore, we discuss ISAAC's operation in brief.

ISAAC restricts the design to limited ADCs per In-Memory Accelerator (IMA). ISSAC also uses 2-bit ReRAM cells and supplies single-bit input voltages to the DAC one at a time in a sequential manner. This allows ISAAC to use low-precision ADCs. Although memristor cells can be designed to store multi-bit values, limiting the number of states in the cell is preferable to ensure stability and reliability. A large number of bits per cell will create a current summation that may require a high-precision ADC to translate it, a non-trivial circuit. Therefore, ISAAC uses four states (two bits) per cell. In addition, the bits of the values are spread across many cells. For instance, in a $128 \times 128$ crossbar, we can use eight consecutive cells in a row to store a 16b value. Therefore, the crossbar will be able to store a $128 \times 16$ matrix. This allows using low-precision and lower-cost ADCs to perform analog to digital conversion of the bit-line currents. However, after the conversion, some shift and add circuitry is required to combine the results into the actual output value. To further reduce the hardware cost for ADCs, ISAAC limits their number and shares among all the crossbars in the IMA. 



Figure \ref{fig:issac}a shows a simplified ReRAM accelerator architecture similar to ISAAC. The accelerator comprises of several chips. Each chip contains many tiles inside it. Each of the tiles host many In-Memory Accelerators (IMA), which are the actual processing engines that performs matrix multiplication and other related operations specific to the application. Each IMA contains several crossbars, ADCs, DACs, S$\&$H, Shift and Add circuitry. 

The crossbars are first programmed with the weights. During inference operation, the inputs are routed to specific IMAs using on-chip network. IMAs then perform dot product operation using the crossbars. After a read cycle, the results appear in the bitlines in form of currents. A Sample \& Hold (S\&H) circuit holds the current until they are supplied to the ADC. ISAAC uses 1.28 giga-samples per second ADCs to convert 128 bitline currents sampled by the S\&H circuit. Since several ADCs are shared among multiple crossbars, at least one ADC needs to become available to start the conversion. As soon as one ADC becomes available, it starts converting each of the bitline currents one by one. Once the conversion begins, the crossbar starts the next read phase. The outputs from the ADCs are accumulated using a Shift \& Add circuit to recover the actual multiplication result. ISAAC tiles also contain CNN-specific circuitry to perform the sigmoid and maxpooling operations. With such a parallel architecture, a large number of dot product operations can be distributed among many IMAs to achieve accelerated CNN inference operation. 

\subsection{ReRAM Failure Model}

Resistive memories such as PCM and ReRAM suffer from soft errors due to changes in resistance, hard errors due to cell imperfection, and wear-out\cite{cong_impact,fault1,fault2,fault3,fault4}. In particular, if ReRAM cells are arranged in a crossbar, there can be two primary categories of failures\cite{cong_impact}. The first one is due to structural reasons caused by the unique organization of the crossbar. These errors are generated from voltage drops, sneak-path currents, and the data values stored in the cells. Such errors can be mitigated with a careful design of the crossbar architecture\cite{cong_xbar}. The second category of failures comes from the ReRAM cells and can be permanent failures or transient failures. Transient failures or soft errors are more challenging to tackle due to their random and non-deterministic nature. The primary cause of soft errors in ReRAM cells is retention failure. Unlike PCM cells, where the resistance drifts over time, retention failure in ReRAM cells is caused by a sudden and abrupt drop or rise of the resistance. The resistance can suddenly drop from High Resistance State (HRS) to Low Resistance State (LRS) or the opposite\cite{fault1,fault2,fault3}. 
\section{Motivation}
\label{sec:motivation}

While errors in regular memory devices are already problematic, their impact can be even more serious when emerging memory technologies are used to process data inside memory in an accelerated fashion. Additionally, traditional error detection and correction methods of guaranteeing reliable memory operation will not be practical in such architectures. Regular ECC for general memory, which detects and corrects errors in conventional memory systems, is less suitable for PIM  devices. It is impractical to invoke a centralized ECC hardware before every in-memory operation. If we want to have faster ECC operation in this scenario, we can have many redundant ECC logic in each of the Processing Engines (PEs) or IMAs to avoid stagnation. However, this will significantly impact power, performance, and cost-effectiveness. A performance-friendly and cost-effective solution can be infrequent memory scrubbing. However, in this scenario, memory scrubbing will have compromised reliability. Between two scrub events, errors may silently propagate through computed faulty results. Additionally, none of the ECC options can detect operation-level failures (e.g., faulty ADC). 


\begin{figure}[h]

        \centering
	\includegraphics[width=\columnwidth]{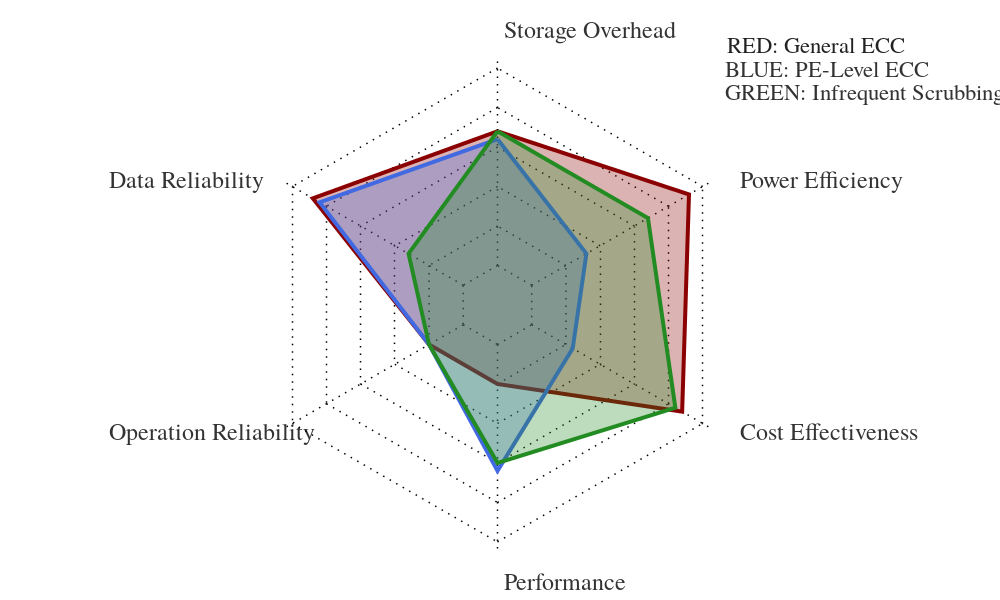}
		\caption{Comparison among reliability options for in-memory computing.}
	
	 \label{fig:motiv}
\end{figure}

Figure \ref{fig:motiv} illustrates the trade-off among competing factors, such as reliability, overhead, power, and performance. Further descriptions of these alternative approaches are provided in Section 4. Accordingly, current PIM architectures lack an appropriate reliability mechanism to guarantee reliability in a deterministic manner. While the relaxed reliability may not be an issue in some applications, the consequence can be devastating in other safety-critical and scientific applications. Special-purpose SoCs that can perform such operations are becoming increasingly popular in the automotive industry and real-time systems. Relaxed reliability is extremely dangerous in many cases, such as autonomous driving and biomedical devices. Therefore, FAT-PIM aims to provide a reliable in-memory operation for ReRAM crossbar-based PIM accelerators while ensuring high data and operation reliability, low storage overhead, high performance, and low hardware complexity.

\section{Design}
\label{sec:design}

This section explores the design space of the FAT-PIM for low-overhead and error-free operation. Initially, we discuss the alternative design options to enable error-tolerant PIM architecture and their associated complexities. Later, the proposed design is discussed.

\subsection{Design Space for Reliable PIM Architecture}

As discussed earlier, we must ensure two things to ensure reliable PIM architectures: (a) error-free operands and (b) the reliability of the operation itself. Since the memory is now responsible for both storing data and performing computation over it, generally, it requires multiple levels of protection. The desired reliability can be achieved through the following straightforward methods.

\subsubsection{Protecting Data from Errors}


\noindent
\textbf{General Purpose ECC:}
General computing systems in both conventional processors and accelerators rely on error detection and correction logic inside the memory controller to check for errors when the data values are fetched from memory. Error Correction Codes (ECC) is calculated and stored alongside the data. Such ECC gives the system the ability to check and correct errors before the operation. {However, such an approach is not efficiently feasible in PIM architectures as the data does not pass through the memory controller as frequently as it does in conventional CPU and GPU architectures.} If we design a PIM system that sends the data to the memory controller before every operation, the system will behave similarly to a conventional CPU, and there will be no benefits of using in-memory computation. In other words, this would be an impractical and naive implementation that is hugely inefficient in exploiting the memory level parallelism.

%

One possible way to prevent stagnation due to centralized error correction logic is to distribute them in different chips or different tiles. Such a method of error correction is also available in conventional memory architecture, where multiple ECC logic is incorporated inside the memory to perform error checks at a lower level. In such In-Memory ECC, errors are checked at a finer granularity, e.g., 128b in DDR5\cite{DDR5Micron}. This ensures faster error checking with a low hardware cost for ECC logic. However, unlike conventional memories, where only a single memory block requires correction at any point in time, PIM architectures require thousands of data values to be checked for errors. For instance, a $128 \times 128$ crossbar with 2-bit cell stores 4kB data alone that need to be accessed during every operation. Additionally, if we take inputs into consideration, we may encounter significant delay before all necessary data in all crossbars are error-checked. Therefore, such distributed error correction logic will still incur significant overhead.

\begin{figure}[h]

        \centering
	\includegraphics[height=9cm,width=\columnwidth]{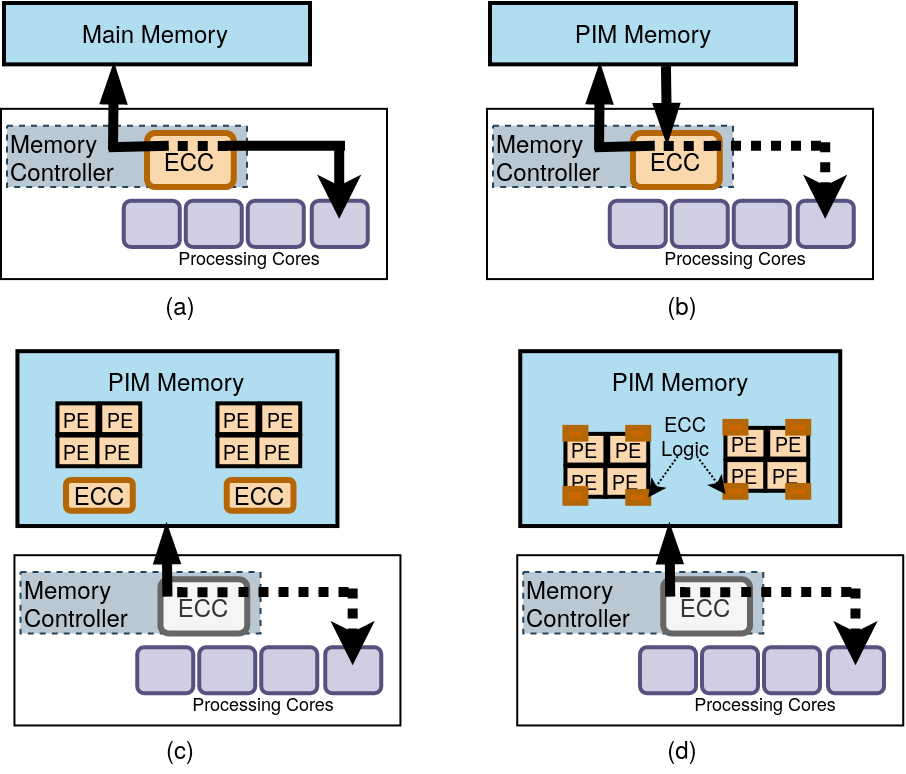}
		\caption{(a) General ECC operation in conventional memory system (b) Stagnation of PIM in centralized ECC (c) Distributed in-memory ECC (d) Aggressive ECC login in all in-memory compute engines}
		
	 \label{fig:eccoptions}
	 \vspace{-1em}
\end{figure}

%

The only way to alleviate this performance bottleneck is to add ECC logic in each low-level compute engine (e.g., the crossbar in this case). To check errors over the required data (weights and inputs), we either need an extensive detection and correction logic or use a smaller error correction logic where the data are fed as smaller chunks sequentially. In both cases, this will tremendously worsen the device's power consumption and add extremely high hardware and performance overheads to the system. Figure \ref{fig:eccoptions} shows different ECC options for such PIM accelerators in comparison with conventional system. Figure \ref{fig:eccoptions}a shows a conventional memory system where data are read from the memory and error-checked in the memory controller before passing them to the core. Figure \ref{fig:eccoptions}b shows a naive PIM architecture where all data are sent to central ECC logic before every operation. Figure \ref{fig:eccoptions}c shows where multiple ECC blocks are used inside the memory. Finally, figure \ref{fig:eccoptions}d shows the case where ECC is added in each PE inside the PIM architecture to avoid ECC bottlenecks.

\noindent
\textbf{Memory Scrubbing:}
One way to keep errors in check in the error-prone memories is to perform periodic scrubbing. Such a method is commonly used in PCM-based memory\cite{PCMScrub}. If we relax the need for error-checking before every operation, we can significantly reduce the hardware cost by periodically performing frequent memory scrubbing. A central scrubber may stall the PIM operations and scrub the weights as well as the eDRAM contents. We can assume that the stored contents (weights and inputs) have pre-calculated ECCs that the scrubber can use to check and correct. However, this scrubbing is limited only to pre-stored data. Newly computed data that need to be re-applied as input will have no ECC and hence cannot be scrubbed. Computing new ECC over the calculated data allows for errors to propagate further. For instance, if soft errors appear within the stored data before the scrubbing operation begins, the system may silently propagate errors through the computed values. While infrequent scrubbing is comparatively low-overhead and hardware simplicity is improved, the reliability is significantly relaxed.

\subsubsection{Reliable Operation}

The second requirement for error-free PIM operation in the ReRAM crossbar accelerator is to guarantee reliable dot product operation. Even if the contents in the crossbar and the input values are checked before the in-memory operation using ECC, external interference and circuit-level glitches can produce a wrong result. Since existing ECC algorithms do not have homomorphic properties, there will be no option to check for and correct errors once the computation is complete. This is a fundamental limitation of in-memory computation and cannot be solved even with the most aggressive ECC circuitry discussed above.

The only existing way to ensure reliable operation is to perform redundant operations. For instance, we can use Triple Modular Redundancy (TMR) to compute the same operation in three different crossbars. We can finally compare and ignore results that do not agree with the majority. While redundant computation will ensure that errors are not silently propagated, it has extremely high hardware costs. There will be a 200\% waste of hardware resources in TMR, which could otherwise be used to compute and improve the throughput.

As discussed, none of the current solutions offer reliability in PIM architecture without significant performance or hardware cost. Therefore, in some aspects, the design options are inherently limited to high-cost methods. Therefore, in the following, we explain our novel solution that can ensure reliability for both the inputs and the operation with minimal hardware and performance cost. 
\subsection{Basic Architecture of FAT-PIM}

\begin{figure}[h]

    \centering
	    \includegraphics[width=\columnwidth]{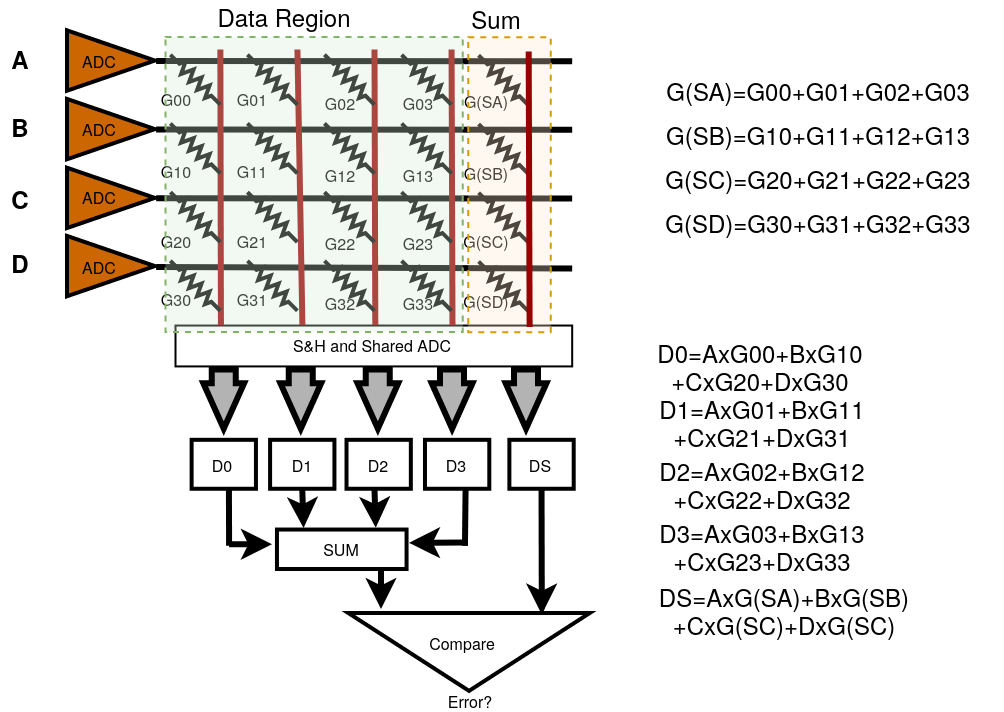}
		\caption{Using summation as homomorphic ECC}
	 \label{fig:sumerror}
\end{figure}

FAT-PIM leverages the fact that memristor crossbar-based PIM accelerators perform current summation on the bit lines in every operation, which can be used to verify the correctness of the values and the operation itself. If we use a dedicated bit line containing cells with conductance values equal to the sum of the conductance of the cells in the corresponding word line, we shall be able to get the summation of all products in that bit line. Figure \ref{fig:sumerror} shows such a crossbar. The conductance values corresponding to the weights (G\textsubscript{00} to G\textsubscript{33}) are stored in the data region of the crossbar. The cells in the additional bit-line hold the summation of the conductance values stored in the cells in each word line. For instance, G(SA) is the summation of G00 to G03. When inputs A, B, C, and D are applied to the crossbar through the word lines, currents will flow through all the bit lines, and the sampled values will represent their corresponding dot products. Also, the value sampled in the summation line will be equal to the summation of values sampled in the rest of the bit lines $D\textsubscript{S} = \sum D\textsubscript{i})$. The operation steps are illustrated in detail in the following discussion.

\noindent
\textbf{Step 1:} The summation values for each word line are pre-calculated before storing the weights. For instance, the corresponding summations are also supplied whenever the PIM device is programmed with values in the crossbar (e.g., weights). The summation calculation can be done at the software level during data preparation or at the hardware level using dedicated adders. Note that we assume that the PIM device will be mostly used for inference purposes, and the values stored in the crossbar will not change unless they are re-programmed.

\noindent
\textbf{Step 2:} The inputs are applied as an analog voltage through the word lines. If used for image recognition hardware, the inputs will belong to either the image that needs to be inferred or the output from another network (CNN or DNN) layer that has been computed previously.

\noindent
\textbf{Step 3:} Currents will flow through the ReRAM cells and accumulate in each bit line. The summation of the currents in each bit line will be sampled using a Sample \& Hold circuitry and will be digitized with the help of an analog to digital circuitry (ADC).

\noindent
\textbf{Step 4:}
The output from the ADCs will be sent to the Sum Checker logic, where the summation of all values coming from the data region will be compared against the value obtained from the sum region.

\noindent
\textbf{Step 5:} If any mismatch is detected, appropriate action will be taken to achieve error-free operation.

The summations will mismatch if the weights deviate from their original value. Hence, the PIM units will be able to check for any errors present in the crossbar. Additionally, this method will guarantee that any error in the operation itself will not go unnoticed. For instance, any outside interference that changes the analog value or any circuit-level glitches in the ADCs and sampling circuitry will be detected. This way, we can have a homomorphic-like error checking operation with the help of a simple summation operation. Fortunately, this method naturally suits the crossbar-based PIM accelerators and can provide reliable operation with minimal overhead.

\subsection{Analysis of FAT-PIM}

\subsubsection{Notations and Foreground}

Even a correctly implemented procedure to program a ReRAM device with the conductance $G_{ij}$ will be subject to natural drifts during the writing process, and the implemented conductance $\overline{G}_{ij}$ can be modeled as a normal distribution, i.e., $\overline{G}_{ij} \sim \mathcal{N}(G_{ij}, \sigma^2)$. This is not the case of an erroneous operation but just reflects the natural uncertainty in programming ReRAM devices. Similarly, the sum of the weights in every row $GS_i = \sum_{j} G_{ij}$ programmed in the \emph{Sum} ReRAM devices, while obtained offline in an exact manner, will be written with an inherent uncertainty as  $\overline{GS}_i = \mathcal{N}(\sum_{j} G_{ij}, \sigma^2)$.

Now, for $n \times n$ PIM, the sum operation over the \emph{Sum} ReRAM devices is the sum of $n$ normally distributed random variables. Hence, with inputs $A_i$, the implemented sum $\overline{DS} = \sum_i A_i \overline{GS}_i$ is distributed as $\mathcal{N}(\sum_i \sum_{j} A_i G_{ij}, \sum_i A_i^2 \sigma^2)$. If the inputs are normalized between 0 and 1 as is common in machine learning, the inputs $A_i$ are upper bounded by 1; hence, the sum of the \emph{Sum} ReRAM devices has more uncertainty than each of the devices themselves but the growth in the uncertainty or standard deviation is $\mathcal{O}(\sqrt{n})$.

Similarly, with the inputs $A_i$, the weighted sum of each of the columns $\overline{D}_j = \sum_i A_i \overline{G}_{ij}$ is a weighted sum of $n$ normally distributed random variables, and is hence given by $\overline{D}_j \sim \mathcal{N}(\sum_i A_i G_{ij}, \sum A_i^2 \sigma^2)$. A sum $S$ of all these $n$ partial sums $\sum_j \overline{D}_j$ for a $n \times n $ PIM is normally distributed, i.e., $\overline{D} \sim \mathcal{N}(\sum_j \sum_i A_i G_{ij}, \sum n A_i^2 \sigma^2 )$. The uncertainty due to noise in this sum $\overline{D}$ grows linearly with the dimension of the crossbar, i.e., the growth is $\mathcal{O}(n)$. 

\subsubsection{Theoretical Result}

Small variances in the writing of $G_{ij}$ are expected and need to be tolerated. Hence, we will ignore small errors in our comparator and flag an error in the operation if and only if the two values are different by higher than a threshold $\delta$. 

\newtheorem{theorem}{\textbf{Lemma}}

\begin{theorem}
Given an acceptable error threshold of $\delta$ and a variance of $\sigma^2$ in operations under normal circumstances, the FAT-PIM will detect any error with $99.9999998\%$ probability for any $n \times n$ crossbar of size $n < \frac{\delta}{12 \sigma}$. 
\end{theorem}
\begin{proof}

Since the sum of the row-wise \emph{Sum} devices $\overline{DS} = \sum_i A_i \overline{GS}_i$ is distributed as $\mathcal{N}(\sum_i \sum_{j} A_i G_{ij}, \sum_i A_i^2 \sigma^2)$, we know the following with the aforementioned probability:

\begin{equation}
   \overline{DS} =  \sum_i \sum_{j} A_i G_{ij} \pm 6 {\left( \sum_i A_i^2 \sigma^2 \right)}^{1/2}
\end{equation}

Similarly, since the sum of the weighted sums of the columns $\overline{D}$ is distributed as $\overline{D} \sim \mathcal{N}(\sum_j \sum_i A_i G_{ij}, \sum n A_i^2 \sigma^2 )$, we know the following with the aforementioned probability:

\begin{equation}
   \overline{D} =  \sum_i \sum_{j} A_i G_{ij} \pm 6 {\left( \sum n A_i^2 \sigma^2 \right)}^{1/2}
\end{equation}

Since we tolerate an acceptance threshold of $|\overline{D} - \overline{DS}| = \delta$, we get the following:
\begin{equation}
   \delta  \geq  6 {\left( \sum A_i^2 \sigma^2 \right)}^{1/2}   + 6 {\left( \sum n A_i^2 \sigma^2 \right)}^{1/2}
\end{equation}
The latter holds true when $\delta \geq 12 n \sigma$; i.e., $n \leq \frac{\delta}{12 \sigma}$.
\end{proof}

\subsubsection{Exposition of Theoretical Results}

Our theoretical analysis shows that the size of the crossbar is a function of the accuracy with which we seek to detect errors. Consider a ReRAM device with a high-resistance state of $0.5 \times 10^{-3}$ S, a low-resistance state of $10^{-3}$ S, that is used  as a 1-bit  device, and an acceptable error tolerance $\delta$ of $0.5 \times 10^{-3}$ S during normal operations. Assuming a standard deviation of $10^{-9}$ S, our results show that our approach will work well with crossbars of size $\frac{0.5 \times 10^{-3}}{12 \times 10^{-9}}$; i.e., PIM with $41,666 \times 41,666$ rows and columns. Later, we also evaluate the likelihood of missed detection due to run-time multi-bit errors after discussing the details of our design.


\subsection{Challenges and Solutions}

While simple, several challenges to practically implementing this in hardware may exist. In the following, we discuss these challenges and how to overcome them.

\subsubsection{Calculation of wordline summation}

The word line summation can be calculated in various ways. The simplest method is to use a software-level extension to pre-calculate the summations before storing them in the crossbar. However, this will add some complexity to the data preparation stage. We calculate the word line sums within the Tile to simplify the software overhead when the weights are programmed into the crossbar. During the programming phase, the values are first stored in the eDRAM buffer, then moved to their destination crossbar. A preparator circuit uses adders to generate the summations before storing them into the crossbar.

\subsubsection{Cell and ADC precision}

We use two bits (four states) per cell ($m=2$), similar to ISSAC. All data (inputs and weights) are first stored in the eDRAM buffer. The preparator circuitry samples and aligns the data adequately based on the programming and inference stage requirements. During the programming stage, apart from sampling each of the values into $n$ $m$-bit values, the preparator circuitry also appends the word line values by appropriate summation. Finally, the sum value is distributed into multiple two-bit values and appended to the word line values.

For a $128 \times 128$ crossbar, if the values of the weights are of $k$ bits each, we can have $v = \frac{m \times 128}{k}$ values in each word lines. Therefore, the maximum value for the summation will be $b = log\textsubscript{2}(v \times 2\textsuperscript{k})$ bits long, requiring $\frac{b}{m}$ additional cells per word line. For instance, if we use 16-bit values in a two-bit cell crossbar architecture, we need ten other cells for each 128-bit word line, requiring only 7.8\% additional storage. The storage overhead for summation increases if the weights need to have more bits. For instance, if we need to design the architecture for 32-bit values, we need 13.7\% additional storage assuming other parameters remain the same. However, the storage overhead can also be reduced by increasing the cell's precision and width of the crossbar. For instance, a three-bit cell crossbar will have 4.1\% storage overhead. To reduce the storage overhead further and have low-overhead sum checking, we calculate the sum over smaller values instead of full 16-bit values. For instance, the preparator circuit will distribute bits in different cells as two-bit values. If we calculate the sum over these two-bit values, the summation will be much smaller and consume less storage space. This method also has some performance benefits, which are discussed later.

The precision of the ADCs depends on the height of the crossbar and the bits per memristor cell. Similar to prior works \cite{ISSAC}, we also allow input bits to be applied as a one-bit voltage per cycle, which reduces the ADC requirement to only nine bits. This can be further reduced to eight bits by flipping the values of the weights if they have a large collective value. FAT-PIM carefully assigns value to additional cells that store the word line summation to use such optimizations for ADC. Since we spread the sum over many two-bit values, the same ADC can be utilized in our case.

\subsubsection{Impact of additional circuitry on performance}

Compared to a conventional memristor-based PIM architecture similar to prior studies\cite{ISSAC,PipeLayer}, the additional primary circuitry is the summation calculation circuitry that calculates the sum of the inputs. When the operation finishes and the currents are available in analog form, they are first sampled and converted to nine-bit digital values. Each of these nine-bit digital values of the current represents the dot product between the inputs and the values stored in the cells on this column. A Shift And Add circuitry is then used to combine them into actual multiplication values. This Shift And Add circuitry is not needed if we store all bits of a value in a single cell. However, using a high number of bits per cell is not practical, and hence the $S\&A$ circuitry is needed anyway. Now, if we calculate the summation over the original value (e.g., 16-bit values), the $S\&A$ circuit first needs to finish combining the results before we can start calculating the sum over all the values. This will have several additional stages of the pipeline beyond the ADC and $S\&A$ circuitry. For instance, if we have 16-bit values, we will have a 39-bit result corresponding to column of the matrix (comprising of eight bit lines) after ADC and $S\&A$ circuitry finishes their operation. There will be 16 such values over which we need to calculate the sum. Therefore, the sum check circuit will have to wait until $S\&A$ circuit finishes preparing a 39-bit value. This will be followed by few cycles to calculate the final summation value.

\begin{figure}[h]

        \centering
	\includegraphics[width=\columnwidth]{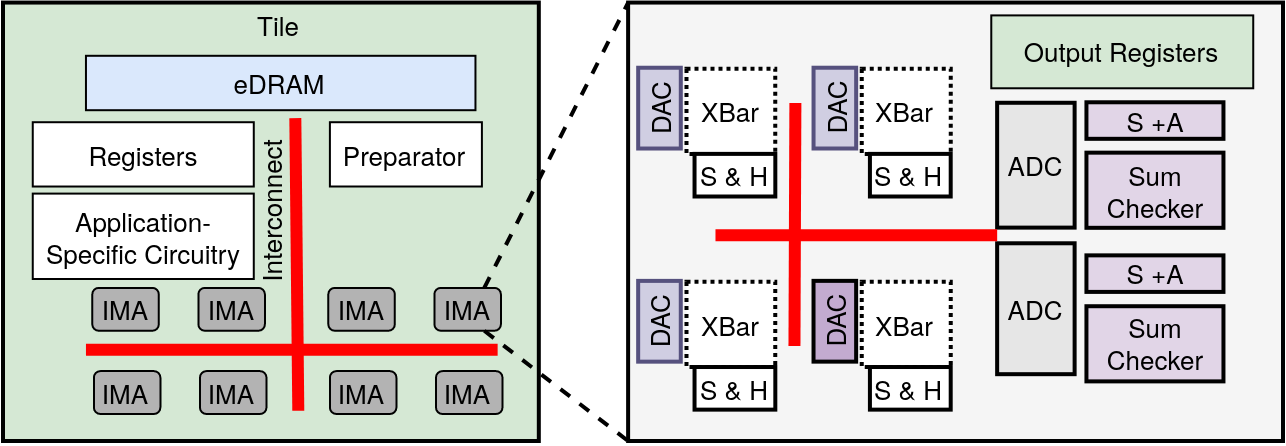}
		\caption{FAT-PIM Architecture}
		
	 \label{fig:fullarch}
\end{figure}

Another way to do this is to implement the sum check circuitry immediately after the ADCs so that the $S\&A$ circuit and the sum check both can work in parallel. Due to the high hardware overheads associated with ADCs, generally, they are limited in number and shared by multiple crossbars in each IMAs. For instance, we can have four ADCs per IMA which are shared by tens of crossbars. After the read operation finishes, the Sample \& Hold $(S \& H)$ circuit samples the analog value in each bit line and holds them to supply it to the ADC. As soon as one ADC becomes available, it starts converting these analog currents one by one in a pipelined fashion, while the crossbar starts to compute the next set of values. Therefore, the ADC outputs a nine-bit value in each cycle. To avoid the high performance and hardware cost of adders, we can simply have one adder that keeps adding the ADC outputs in a pipelined manner. The $S \& H$ circuit also operates in parallel, and hence there is no additional performance overhead if the summation is done in this way. For a $128 \times 128$ crossbar, the ADC circuitry will finish converting all bit line currents in 128 cycles, and the following cycle will have the final sum as well as the last value from the $S \& A$ circuit. The comparator then compares two sums that take only one cycle. There are also additional cycles needed in the ADC to convert the bit lines corresponding to the stored summation.

However, now since the summation is performed over the non-accumulated values collected from each bit line, we also need to prepare the stored sum value accordingly. Now, we need to calculate 128 two-bit values to get the sum in a $128\times128$ crossbar during the program phase. Fortunately, this optimization has a positive impact on storage overhead since we are now calculating summation over smaller values. This will also essentially reduce the storage overhead significantly, from 7.8\% to only 3.9\%.

\subsubsection{Input Errors}

A mismatch between the summation calculated over the bit line values and the stored summation will capture faults in the crossbar cells, ADC, S\&H, and other circuitry in the path between the crossbar and the sum check circuit. However, errors can also happen in the input bits supplied to the crossbar. The inputs are stored in the eDRAM buffer and supplied to the crossbar by the preparator during regular operation. Since the inputs are handled as regular memory data, they can be protected using regular ECC. Therefore, any data stored in the eDRAM buffer can also have its ECC stored alongside it. The preparator controls the flow of the inputs from the eDRAM to the IMAs and can check for errors using the ECC. Therefore, ensuring the correctness of the input values before applying them to the crossbar.

\subsection{Putting It Altogether}

Figure \ref{fig:fullarch} shows the overall architecture of our error-tolerant PIM architecture for the memristor crossbar. Each Tile contains an eDRAM buffer (to temporarily store data), preparator circuitry, registers, and several IMAs. Tiles can also have some application-dependent circuitry specific to certain operations. For instance, CNN operation requires a sigmoid and max-pooling operation, which can be handled by such application-specific hardware. The preparator circuitry aligns the bits properly before sending them to the crossbar and checks for errors using ECC before sending them to the IMAs. Tiles also hold multi-purpose registers that control the operations, modes, and also temporarily holds the data. 

During regular operation, the preparator circuitry sends the input values in a digital format which are later converted to analog voltages and applied to the target crossbar. The crossbar will generate dot product results in the form of analog currents, which are sampled in the S\&H circuit. Later, the analog currents are routed to an available ADC, which starts converting the bit-line currents one by one. The shift-and-add circuit and sum calculator works in parallel over the outputs generated from the ADC and finally generate both the result in original format and the corresponding summation of values. Finally, the calculated summation and the summation calculated by the crossbar is compared to identify any faults in operation. 

\subsection{Error Correction}

While the summation checker will be able to identify operation failures and errors within the crossbar cell, it does not have inherent correction capability. Therefore, FAT-PIM will only monitor for errors inside the crossbar and interrupt the operation once an error is identified. In the event of error detection, the IMA signals an error to the Tile instead of sending the computed value. The Tile then takes appropriate actions such as re-programming the crossbar or raising an interrupt that is sent to the host to take appropriate action. To ensure correct operation after an error is detected, we stall the IMA and re-program the cells from the values stored in the eDRAM buffer. If errors are detected multiple times even after re-programming, we can conclude that a permanent fault has occurred in one of the crossbar components, and hence it cannot be used further.

\subsection{Likelihood of Missed Detection}

The crossbar will compute vector summation ($S\textsubscript{BL}$) over each of the bit lines. These summation values are then added together into a bigger sum ($\sum S\textsubscript{BL}$) that is compared against the collective summation ($\sum S\textsubscript{WL}$)  of all worldline summations ($S \textsubscript{WL}$). $S \textsubscript{WL}$ values are pre-calculated and stored when the crossbar is programmed. Here we discuss the likelihood of an event where two distinct bit flips in the crossbar cells cause a detection failure because of equal $\sum S\textsubscript{BL}$ and $\sum S\textsubscript{WL}$. Such an event can be possible in two general cases. First, there can be multiple errors in either the data region or the summation region of the crossbar, which results in unchanged $\sum S\textsubscript{BL}$ or $\sum S\textsubscript{WL}$. The second case is where both $\sum S\textsubscript{BL}$ and $\sum S\textsubscript{WL}$ change to erroneous $\sum S\textsubscript{BL}\textsuperscript{*}$ or $\sum S\textsubscript{WL}\textsuperscript{*}$, respectively, but they remain equal. 


\begin{figure}[h]

        \centering
	\includegraphics[width=\columnwidth]{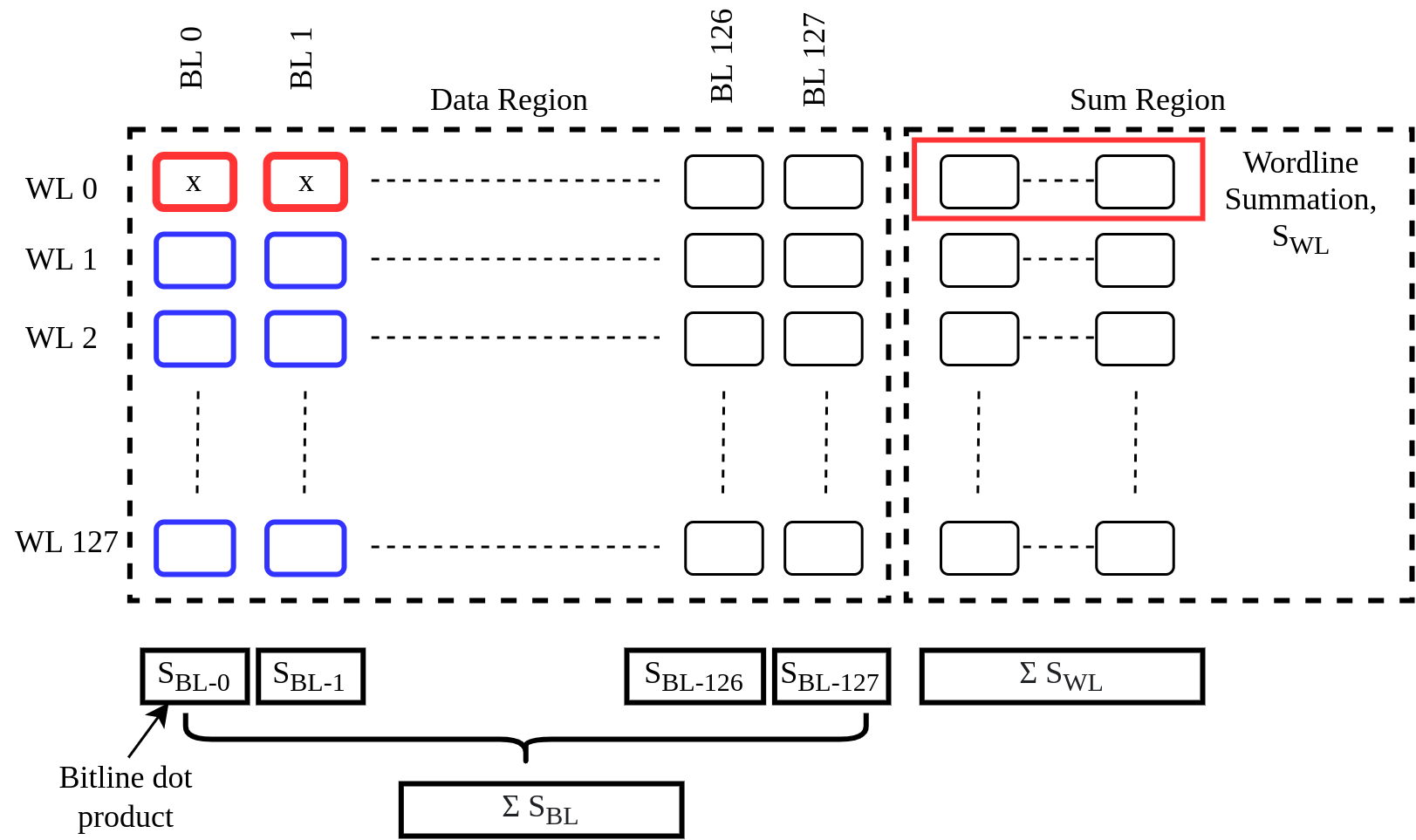}
		\caption{Tolerating multiple errors in crossbar}
		
	 \label{fig:errorcase}
\end{figure}

\begin{table}[]
\scriptsize
\centering
\label{tab:errorrate}
\caption{Experimentally calculated probability of missed detection due to multi-bit error}
\begin{tabular}{|l|l|l|l|}
\hline
                   & \textbf{64x64 crossbar} & \textbf{128x128 crossbar} & \textbf{512x512 crossbar} \\ \hline
\textbf{16b input} & 1.25E-11                & 5.3E-12                   & 1.9E-12                   \\ \hline
\textbf{8b input}  & 1.9E-11                 & 1.06E-11                  & 7.8E-12                   \\ \hline
\end{tabular}
\end{table}

For the first case, both errors occur in two different locations within either the data region or the summation region. In this case, they may fall along the same bit line or across different bit lines. For two random changes in two different cells within a bit-line, the likelihood of generating a same $S\textsubscript{BL}$ can be roughly estimated to $\frac{1}{total\ number\ of\ possible\ sum\ values}$, or $\frac{1}{(2\textsuperscript{m}-1)\times w}$, for m-bit ReRAM cell and $w$ wordlines/columns in the crossbar. The complete multiplication operation finishes when all input bits are applied through the DAC. For $i$-bit input value, $S\textsubscript{BL}$ will remain unchanged only if the combination of input bits corresponding to the position of the faulty cell retains the same for the complete $i$ iterations. For instance, if the input bit from DAC changes from 1 to 0 for one of the word lines where the error happened, $S\textsubscript{BL}$ will no longer remain the same, and the error will be detected. For N randomly chosen faulty cells at different word line positions, the likelihood that the input pattern across the DAC will remain identical for the entire $i$-bit input is $(\frac{1}{2\textsuperscript{$N \times i$}})$. Therefore, if the probability of a random single-bit error is $p$, the likelihood of another occurrence that causes similar summation will be $p^2 \times \frac{1}{(2\textsuperscript{m}-1)\times w} \times (\frac{1}{2\textsuperscript{$N \times i$}}) $, or $p^2 \times p^*$, where $p^*$ is the likelihood of failed detection should a multi-bit error occurs. If the errors share the same wordline, the bitline summations in each of the bitlines will change. Now, the probability of $\sum S\textsubscript{BL}$ remaining unchanged can be similarly calculated to $p^2 \times \frac{1}{(2\textsuperscript{s}-1)\times w} \times (\frac{1}{2\textsuperscript{$N \times i$}}) $, where s is the size of bitline summation (e.g., 9-bit for 128x128 crossbar). For the second case where both summation values ($\sum SBL$  and $\sum SWL$) change, the likelihood of one of them accidentally being similar to the other can also be represented as $p^2 \times \frac{1}{(2\textsuperscript{s}-1)\times w} \times (\frac{1}{2\textsuperscript{$N \times i$}}) $ since they each can have $(2\textsuperscript{s}-1)\times w$ possible values. To better understand the multi-bit error scenario and consider all possible cases that the above theoretical description may not cover, we also conduct experimental analysis by randomly injecting multi-bit errors into a single crossbar. For all cases, the probability of correct detection was close to 1. Considering the worst-case probability of error (1E-3) in restive memory devices reported in prior studies\cite{sills2015high,zhang2018exploring}, the experimentally calculated probability of missed detection in FATPIM for two-bit errors is tabulated in Table \ref{tab:errorrate}.

\section{Methodology}
\label{sec:methodology}

We have used an in-house simulator to model and evaluate the performance of the FAT-PIM design. We have adopted the design parameters from ISSAC architecture and additionally incorporated FAT-PIM's reliability logic. Table \ref{table:eval} lists the parameters we have used in our evaluation. We simulate a ReRAM crossbar-based PIM architecture with eight chips, 16 tiles per chip, 12 In Memory Accelerators (IMA) per tile, and 12 crossbars per IMA. We manually program the crossbars with random values for weights before evaluation. Pre-computed random inputs are also stored in the eDRAM buffer that feeds them to the crossbar during in-memory operation. 

\begin{table}[]

\scriptsize
\centering
\caption{FAT-PIM's Simulation Parameters}
\vspace{2em}
\begin{tabular}{|lll|}
\hline
\multicolumn{3}{|l|}{\textbf{Simulation Parameters}}                                                                            \\ \hline
\multicolumn{1}{|l|}{\multirow{2}{*}{\textbf{Overall Architecture}}} & \multicolumn{1}{l|}{Number of Chips}     & 8             \\ \cline{2-3} 
\multicolumn{1}{|l|}{}                                               & \multicolumn{1}{l|}{Number of Tiles}     & 16 per chip   \\ \hline
\multicolumn{1}{|l|}{\multirow{2}{*}{\textbf{Tile Parameters}}}      & \multicolumn{1}{l|}{eDRAM}               & 42MB per tile \\ \cline{2-3} 
\multicolumn{1}{|l|}{}                                               & \multicolumn{1}{l|}{Number of IMAs}      & 12 per tile   \\ \hline
\multicolumn{1}{|l|}{\multirow{4}{*}{\textbf{IMA Parameters}}}       & \multicolumn{1}{l|}{Number of ADCs}      & \begin{tabular}[c]{@{}l@{}}4 per IMA\\ 1.28Giga-Samples/sec\end{tabular}     \\ \cline{2-3} 
\multicolumn{1}{|l|}{}                                               & \multicolumn{1}{l|}{Number of S\&A}       & 4 per IMA     \\ \cline{2-3} 
\multicolumn{1}{|l|}{}                                               & \multicolumn{1}{l|}{Sum Checker}         & 4 per IMA     \\ \cline{2-3} 
\multicolumn{1}{|l|}{}                                               & \multicolumn{1}{l|}{XBAR}                & 12 per IMA    \\ \hline
\multicolumn{1}{|l|}{\multirow{4}{*}{\textbf{XBAR Parameters}}}      & \multicolumn{1}{l|}{Row Width}           & 128           \\ \cline{2-3} 
\multicolumn{1}{|l|}{}                                               & \multicolumn{1}{l|}{Column Width (data)} & 128           \\ \cline{2-3} 
\multicolumn{1}{|l|}{}                                               & \multicolumn{1}{l|}{Column Width (sum)}  & 5             \\ \cline{2-3} 
\multicolumn{1}{|l|}{}                                               & \multicolumn{1}{l|}{S\&H}                 & 128 per XBAR  \\ \hline
\multicolumn{1}{|l|}{\multirow{2}{*}{\textbf{Memory Latency\cite{cong_xbar,ISSAC}}}}       & \multicolumn{1}{l|}{Read Latency}        & 100ns         \\ \cline{2-3} 
\multicolumn{1}{|l|}{}                                               & \multicolumn{1}{l|}{Write Latency}       & 200ns         \\ \hline
\end{tabular}
\label{table:eval}
\end{table}

While multiple IMAs can operate in parallel, there are a limited number of ADCs per IMA. Therefore, ADCs are utilized in a shared manner among the crossbars. The values generated by the crossbars are fed to the ADCs, S \& A, and Sum Check circuit in a pipelined fashion. We pre-generate random input traces with varying pipeline delays to evaluate the performance. For example, the application {\tt App\_X\_Y} will have Y cycles delay after every X cycle. This analysis tries to mimic the real-world scenario where it is always not possible to maintain a continuous supply of data to the IMAs due to dependencies and other contentions outside the IMAs. To evaluate the system's reliability, we perform random injections of random and transient faults in the system. Accordingly, we evaluate FAT-PIM's detection capability and correction overhead in the presence of errors. We also perform a sensitivity analysis by varying the ADC throughput and number of summation bit-lines.
\section{Results}
\label{sec:result}

In this section, we present our evaluation results in terms of FAT-PIM's performance and fault tolerance capability.

\subsection{Performance Impact}

FAT-PIM checks for error by re-calculating the summation during every crossbar operation. Although the summation calculation is hidden underneath the ADC and S\&A operation, the ADC circuit must also perform the conversion for the summation bit lines in the crossbar. For example, for a crossbar having 128 bit-lines for data, there will be five additional summation columns that need to be converted using ADC in addition to the data lines. This requires five additional cycles. Note that the additional slowdown can be easily eliminated by using a faster ADC. For instance, if we use ADCs with 1.33 giga-samples capability per second, the performance overheads can be hidden. However, we keep the ADC throughput identical to understand the overheads.

\begin{figure}[h]

        \centering
	\includegraphics[width=\columnwidth]{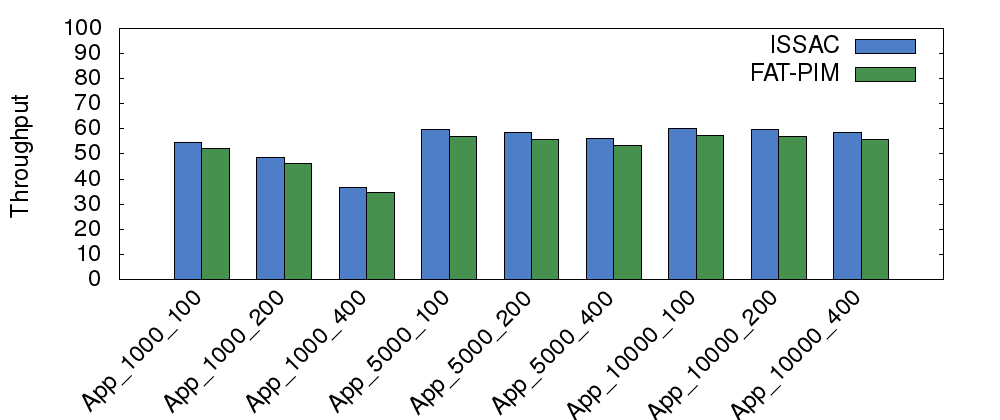}
		
	 \caption{FAT-PIM's Impact on performance.}
	 \label{fig:perfsynth}
\end{figure}

We measure throughput (Figure \ref{fig:perfsynth}) in terms of successful dot product results produced per cycle throughout the accelerator to evaluate the performance. {\tt App\_0\_0} is the ideal scenario for an application that is able to best utilize the hardware as there are always available inputs that can be fed to the IMAs. On the other hand, the throughput decreases with an increase in input delays. With such delays, the pipelined ADCs cannot properly utilize their resource. For instance, {\tt App\_1000\_400} performs worst due to frequent delays and pipeline stalls. After every 1000 cycle, {App\_1000\_400} incurs a pipeline delay for 400 cycles. This causes frequent bubbles in the ADC pipeline and hence reduced performance. Additionally, due to the extra cycles needed for FAT-PIM operation, its throughput is slightly less than the corresponding baseline. On average, FAT-PIM has a 4.9\% additional performance impact over the baseline system.

\subsection{Reliability Analysis: Error Detection}
\label{sec:reliab}
In this section, we discuss FAT-PIM's effectiveness in terms of error detection. To analyze this, we inject random faults to ReRAM cells during run-time operation. A ReRAM crossbar cell with a deviated value will produce an incorrect result which will be detected after summation calculation. During every operation, we compare the summation values and log the outcome. 

The rate of failures in ReRAM cells depends on many parameters, including cell imperfections, the number of bits per cell, and environmental conditions. Generally, multi-bit resistive memory cells are less error-tolerant compared to the single-bit (two states) alternative. Apart from the failures occurring in the ReRAM cell, there can also be other factors that contribute to the incorrect result. For instance, circuit-level glitches may happen in S\&H, ADC, and S\&A circuits producing wrong computation results. Due to many factors involving faulty results, we simplify our evaluation by considering a large range of Failure In Time (FIT) values from studies performed in resistive memory failure analysis \cite{fault1,fault2,fault3,fault4}. For instance, Jubong et al. \cite{fault3} reports MTTF (Mean Time To Failure) for ReRAM cells. Based on the experimental temperature-dependent retention time model reported by Jubong et al., the MTTF of ReRAM cell operating at typical upper-bound of tolerable computing system temperature $(85^{\circ}C)$ can be $2.2 \times 10 ^ 6$ seconds, which can be translated to $1.6 \times 10\textsuperscript{-3}$ FIT per hour per cell. We consider a similar value to represent a real-world scenario and extend the failure rate to 1.6 FIT per hour per cell (equivalent to operating in $\approx160^{\circ}C$ according to the data reported by Jubong et al.) to represent extreme cases. Finally, we analytically estimate the FIT for the device and the corresponding number of total faulty cells for a given time. 

\begin{figure}[h]

        \centering
	\includegraphics[width=\columnwidth]{ 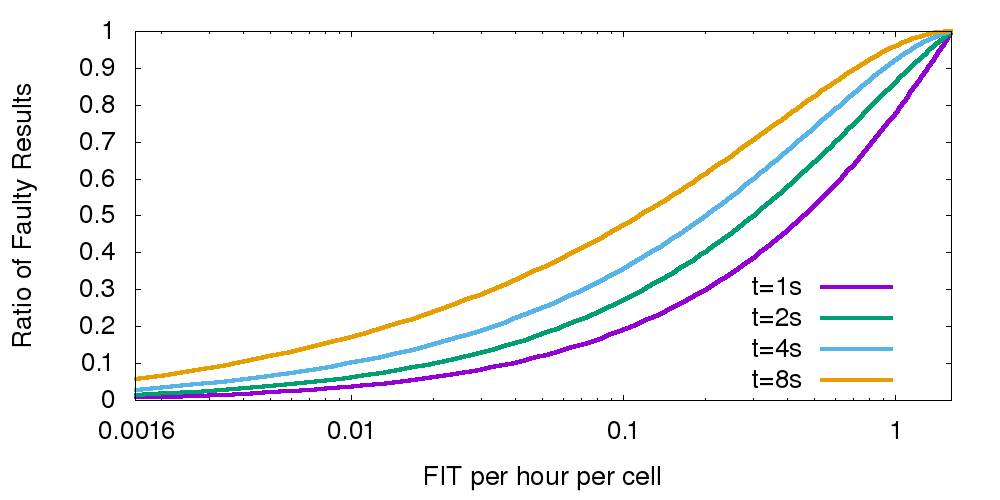}
		
	 \caption{Detection of faulty operation.}
	 \label{fig:errortol}
\end{figure}

To estimate the error detection behavior, we periodically introduce faults into ReRAM cells in the crossbar and try to detect them through summation calculation. Accumulated faults also depend on when operations begin after programming the ReRAM cells. With a longer delay, the likelihood of errors in ReRAM cells increases. With such varying delays, we calculate how many results are detected as faulty by the summation checker. Figure \ref{fig:errortol} shows such data. As shown in the figure, a lower failure rate ($1.6 \times 10\textsuperscript{-3}$  to 0.1) has a percentage of faulty results below 20\%. On the other hand, a higher failure rate (FIT per cell per hour > 1) will have many faulty cells in the system. A high failure rate can easily add at least one faulty cell per crossbar with even distribution of errors. As a result, all computed results are detected as faulty. Besides checking the summation values, we also manually inspect the results to observe any detection failure. Such detection failure may be caused by extreme cases of equal changes of values in two cells in a crossbar in the opposite direction to generate an equal sum. After our simulation, we also compare the results manually with a golden reference having no error and we have encountered no such cases.

\subsection{Error Correction Overhead}

Unlike conventional ECC, which can perform error correction after detecting errors, our scheme can only detect errors using the summation method. Therefore, error correction must be handled separately. While there can be different ways to augment FAT-PIM to allow error correction, here we evaluate a simple and straightforward method of re-programming the crossbar after an error is discovered. We allow a redundant copy of the crossbar's data to reside in the eDRAM buffer. Any data stored in eDRAM (e.g., weights and inputs) will have conventional ECC alongside it to check for errors in the preparator circuit before sending it to the IMA. 

We evaluate this error correction scenario with a random injection of errors based on the failure rate. Once an error is detected, the crossbar stalls and signals the corresponding IMA for error correction. The IMA then allows the Tile to re-program the faulty crossbar by copying data from eDRAM to the crossbar. For the $128 \times 128$ crossbar, 128 consecutive write operation is done to re-program the entire crossbar.


\begin{figure}[h]

        \centering
	\includegraphics[width=\columnwidth]{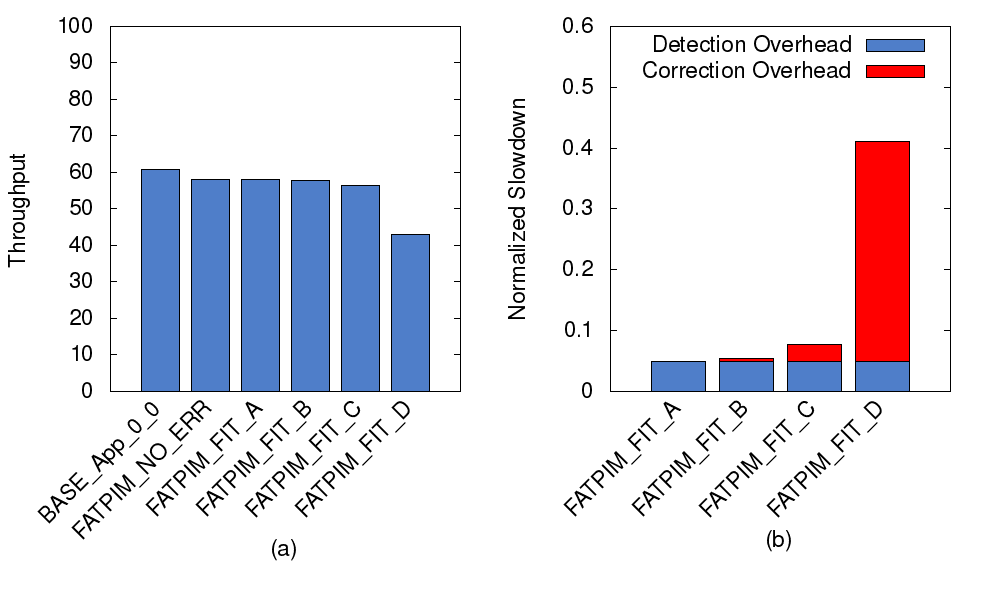}
		\caption{(a) Impact of FAT-PIM's error correction on performance, (b) Breakdown of error detection and correction overhead for soft errors }
		\vspace{-1em}
		
	 \label{fig:errorcor1}
\end{figure}


		

Figure \ref{fig:errorcor1}a and \ref{fig:errorcor1}b show how such correction method can impact the performance. Figure \ref{fig:errorcor1}a shows the impact on throughput, and Figure \ref{fig:errorcor1}b shows the breakdown of correction and detection overheads. {\tt BASE\_App\_0\_0} runs {App\_0\_0} without FAT-PIM. {\tt FATPIM\_NO\_ERR} has only detection overhead since no error is injected. Then we introduce random faults in the crossbar with different failure rates. FIT-A, FIT-B, FIT-C, and FIT-D corresponds to failure rate of $1.6 \times 10 \textsuperscript{-3}$, $1.6 \times 10 \textsuperscript{-2}$, $1.6 \times 10 \textsuperscript{-1}$ and 1.6 failure per hour per cell respectively. Here, as explained in Section \ref{sec:reliab}, A and B are realistic cases, and C and D are extreme cases. As shown in the figure, the correction overhead is extremely minimal and less visible with a relatively low failure rate. However, with a higher failure rate, the system incurs significant stalls due to frequent re-program events. 

\subsection{Sensitivity}

\begin{figure}[h]

        \centering
	\includegraphics[width=\columnwidth]{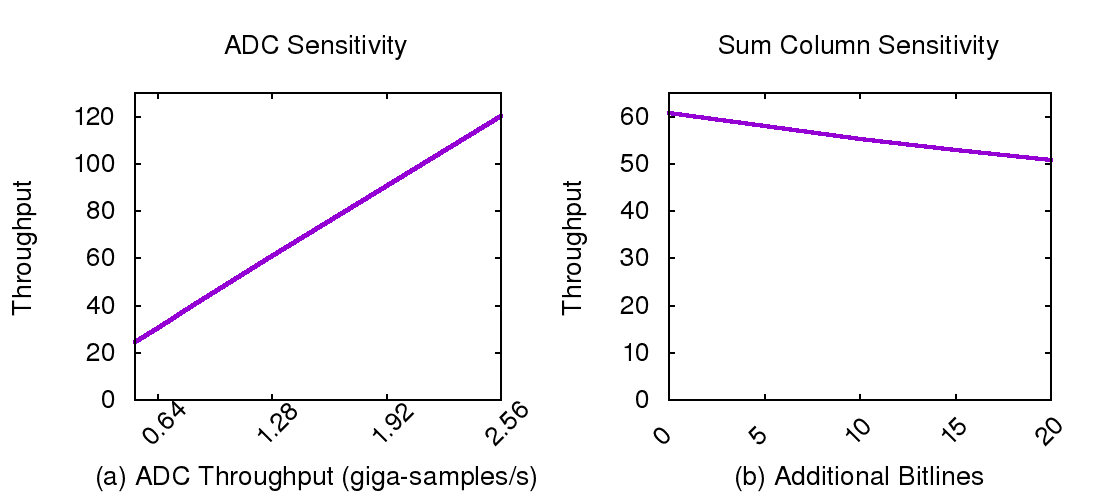}\vspace{-2mm}
		\caption{Sensitivity Analysis}
		\vspace{-2mm}
	 \label{fig:sens}
\end{figure}

Figure \ref{fig:sens} shows the impact on performance after varying the ADC latency and summation conversion overhead. Our design with $128 \times 128$ crossbar and summation calculated over 2-bit cell values uses five additional bit lines to store the summations. With a different crossbar size or different bits-per-cell, this value can vary. Therefore, we evaluate throughput after varying this parameter. Figure \ref{fig:sens}a shows such analysis after varying ADC throughput from 0.52 giga-samples per second to 2.56 giga-samples per second. As expected, the throughput increase with faster ADC. Similarly, in \ref{fig:sens}b, we vary the number of additional bit-lines needed for FAT-PIM. 

\section{Discussion}
\label{sec:discussion}

\subsection{Error tolerance of ML models}

Due to the extensive computation demand for machine learning applications, machine learning operations are generally performed in HPC settings or in general/special purpose accelerators. An operation can be divided into training and inference. During training, the weights and biases of the neural networks are computed based on a training data set. The trained model is then used to perform a prediction during inference. Since such operations are performed in computing resources, such as HPC and accelerators, errors may occur, impacting the outcome. 

Several prior studies \cite{NVIDIA, zhang2019quantifying} have performed extensive studies to understand the impact of hardware-level faults on ML applications. Errors during the training phase have less impact on the outcome than in the inference phase. This is because the training algorithms inherently measure the deviation of the computed outcome with the ideal data to train the network correctly. Therefore, errors in one iteration may be corrected in the next iteration, minimizing the error's impact on the training. However, Zhao et al.\cite{zhang2019quantifying} showed that faults could sometime cause Silent Data Corruption (SDC) which can be problematic. On the other hand, errors during inference have more catastrophic consequences. While some errors may have no impact on the inference and may produce correct prediction results in the presence of errors, some errors can easily cause the network to make a wrong prediction. Prior works have concluded that the resilience of a machine learning model varies with the type of the model, values used, the scale of dependency, and the type and number of layers used in the model. For instance, Guanpeng et al.\cite{NVIDIA} illustrate an experiment where the injection of error causes an autonomous vehicle to classify an incoming transportation truck as a flying bird. Such misprediction can have a devastating consequence as the system will not take proper action to avoid danger. In fact, such misdirection in safety-critical systems is intolerable. For instance, ISO 26262 standard specifically mandates SoC that performs DNN application for the autonomous vehicle to operate with specific reliability guarantee \cite{26262}.


\subsection{Hardware Overhead}

The additional hardware cost for error detection is negligible in FAT-PIM. To estimate the area overhead of the proposed design, we use the data reported in the prior arts for memristor crossbar-based PIM architecture \cite{ISSAC} and adder design in 32nm. Based on our calculation, the FAT-PIM error detection logic roughly has an area overhead of $1.86 \times $10$ \textsuperscript{-2}\%$ Note that we categorize the overhead for additional ReRAM cells as storage overhead, which is only 3.9\%.

\subsection{Tolerating Hard Errors}

Permanent errors occur when a ReRAM cell fails due to endurance failure or manufacturing defects. While ReRAM's endurance is not as good as DRAM, they have slightly better endurance compared to PCM \cite{cong_impact}. Since hard errors or permanent failures are relatively rare compared to soft errors, they can be handled relatively easily. For instance, post-production probing of cells can be used to identify crossbars with faulty cells and can be mapped to redundant crossbars. It is also possible to manufacture the crossbars with redundant rows and columns to isolate faulty cells within crossbars. Such hard errors in the memristor crossbar have been studied in prior literature \cite{cong_impact,chaudhuri2019hardware} and can be implemented alongside FAT-PIM.

\subsection{Reliability For Other PIM Technologies}

Besides memristor crossbar-based PIM devices, there exist other PIM technologies that target different memory technologies. For instance, the in-memory method of computation in DRAM \cite{ambit,gao2019computedram} is fundamentally different from memristor-based PIM devices. Due to such differences, it is challenging to design a single scheme that can work for all devices. Therefore, we believe that different approaches are required for different PIM devices to ensure reliability. For instance, the memristor crossbar performs current summation in the bit line, enabling FAT-PIM to use summation as homomorphic ECC. On the other hand, DRAM in-memory computation performs several different operations such as AND, OR, and NOT, which makes designing a general error detection for both inputs and operations challenging. We leave such exploration for future research.

\section{Related Works}
\label{sec:related}

\subsection{General Purpose PIM Architecture}
\vspace{1em}
In-memory computation has been widely explored for different memory technologies. Ambit\cite{ambit}, ComputeDRAM\cite{gao2019computedram} focuses on operating on data stored in DRAM cells. The authors present the idea that the charge level in capacitors can indicate the result of the operation. It is possible to activate multiple rows and measure the impact on the bit line from the capacitors to determine the bitwise operation result. Such PIM architecture can perform bitwise operations such as AND, OR, and NOT. General-purpose in-memory computation in NVM cells is also explored in the past. Pinatubo \cite{li2016pinatubo} presents ways to use the combined resistance level of the circuit to perform bitwise operations easily in resistive memories. Ensuring error-free operation in general-purpose PIM architecture is also equally challenging due to their diverse nation of operation. We leave such exploration for future research.

\subsection{Special Purpose In-Memory Accelerators}

The special-purpose application of traditional near memory computing architecture that performs certain types of operation in an accelerated fashion is also widely explored in recent literature. Such accelerators are generally special-purpose accelerators built for certain applications such as machine learning. TensorDIMM\cite{kwon2019tensordimm} and DaDianNao\cite{chen2014dadiannao} are examples of such special purpose architecture. These architectures combine processing core and memory storage technology into a single package in a way that can compute a large number of operations in parallel. Such architectures also carry special-purpose circuitry specific to certain operations, such as max-pooling and sigmoid.



\section{Conclusion}

Processing In Memory (PIM) accelerators can tremendously speed up data-intensive applications. However, despite the tremendous potential of PIM devices, their reliability is rarely explored. In this paper, we explore the reliability issues in memristor crossbar-based PIM accelerators and show its fundamental reliability limitation. We propose FAT-PIM, which can make such a system error-tolerant without significant performance overhead and hardware cost. Our evaluation shows that we can improve the error tolerance significantly with only 4.9\% performance cost and 3.9\% storage overhead. 



\bibliographystyle{IEEEtranS}
\bibliography{refs}

\begin{thebibliography}{10}
\providecommand{\url}[1]{#1}
\csname url@samestyle\endcsname
\providecommand{\newblock}{\relax}
\providecommand{\bibinfo}[2]{#2}
\providecommand{\BIBentrySTDinterwordspacing}{\spaceskip=0pt\relax}
\providecommand{\BIBentryALTinterwordstretchfactor}{4}
\providecommand{\BIBentryALTinterwordspacing}{\spaceskip=\fontdimen2\font plus
\BIBentryALTinterwordstretchfactor\fontdimen3\font minus
  \fontdimen4\font\relax}
\providecommand{\BIBforeignlanguage}[2]{{%
\expandafter\ifx\csname l@#1\endcsname\relax
\typeout{** WARNING: IEEEtranS.bst: No hyphenation pattern has been}%
\typeout{** loaded for the language `#1'. Using the pattern for}%
\typeout{** the default language instead.}%
\else
\language=\csname l@#1\endcsname
\fi
#2}}
\providecommand{\BIBdecl}{\relax}
\BIBdecl

\bibitem{DDR5Micron}
``{Introducing Micron® DDR5 SDRAM: More Than a Generational Update},''
  \url{https://bit.ly/37w4ZWb}, accessed: 02/06/2021.

\bibitem{INM_Genome1}
T.~Ahmad, N.~Ahmed, J.~Peltenburg, and Z.~Al-Ars, ``Arrowsam: In-memory
  genomics data processing using apache arrow,'' in \emph{2020 3rd
  International Conference on Computer Applications \& Information Security
  (ICCAIS)}.\hskip 1em plus 0.5em minus 0.4em\relax IEEE, 2020, pp. 1--6.

\bibitem{angizi2019mrima}
S.~Angizi, Z.~He, A.~Awad, and D.~Fan, ``Mrima: An mram-based in-memory
  accelerator,'' \emph{IEEE Transactions on Computer-Aided Design of Integrated
  Circuits and Systems}, vol.~39, no.~5, pp. 1123--1136, 2019.

\bibitem{PCMScrub}
M.~Awasthi, M.~Shevgoor, K.~Sudan, B.~Rajendran, R.~Balasubramonian, and
  V.~Srinivasan, ``Efficient scrub mechanisms for error-prone emerging
  memories,'' in \emph{IEEE International Symposium on High-Performance Comp
  Architecture}, 2012, pp. 1--12.

\bibitem{chaudhuri2019hardware}
A.~Chaudhuri, B.~Yan, Y.~Chen, and K.~Chakrabarty, ``Hardware fault tolerance
  for binary rram crossbars,'' in \emph{2019 IEEE International Test Conference
  (ITC)}.\hskip 1em plus 0.5em minus 0.4em\relax IEEE, 2019, pp. 1--10.

\bibitem{chen2014dadiannao}
Y.~Chen, T.~Luo, S.~Liu, S.~Zhang, L.~He, J.~Wang, L.~Li, T.~Chen, Z.~Xu,
  N.~Sun \emph{et~al.}, ``Dadiannao: A machine-learning supercomputer,'' in
  \emph{2014 47th Annual IEEE/ACM International Symposium on
  Microarchitecture}.\hskip 1em plus 0.5em minus 0.4em\relax IEEE, 2014, pp.
  609--622.

\bibitem{fault1}
B.~Gao, H.~Zhang, B.~Chen, L.~Liu, X.~Liu, R.~Han, J.~Kang, Z.~Fang, H.~Yu,
  B.~Yu \emph{et~al.}, ``Modeling of retention failure behavior in bipolar
  oxide-based resistive switching memory,'' \emph{IEEE Electron Device
  Letters}, vol.~32, no.~3, pp. 276--278, 2011.

\bibitem{gao2019computedram}
F.~Gao, G.~Tziantzioulis, and D.~Wentzlaff, ``Computedram: In-memory compute
  using off-the-shelf drams,'' in \emph{Proceedings of the 52nd annual IEEE/ACM
  international symposium on microarchitecture}, 2019, pp. 100--113.

\bibitem{fault2}
A.~Kawahara, R.~Azuma, Y.~Ikeda, K.~Kawai, Y.~Katoh, Y.~Hayakawa, K.~Tsuji,
  S.~Yoneda, A.~Himeno, K.~Shimakawa \emph{et~al.}, ``An 8 mb multi-layered
  cross-point reram macro with 443 mb/s write throughput,'' \emph{IEEE Journal
  of Solid-State Circuits}, vol.~48, no.~1, pp. 178--185, 2012.

\bibitem{INM_personalized_recom1}
L.~Ke, U.~Gupta, B.~Y. Cho, D.~Brooks, V.~Chandra, U.~Diril, A.~Firoozshahian,
  K.~Hazelwood, B.~Jia, H.-H.~S. Lee \emph{et~al.}, ``Recnmp: Accelerating
  personalized recommendation with near-memory processing,'' in \emph{2020
  ACM/IEEE 47th Annual International Symposium on Computer Architecture
  (ISCA)}.\hskip 1em plus 0.5em minus 0.4em\relax IEEE, 2020, pp. 790--803.

\bibitem{INM_personalized_recom2}
L.~Ke, X.~Zhang, J.~So, J.-G. Lee, S.-H. Kang, S.~Lee, S.~Han, Y.~Cho, J.~H.
  Kim, Y.~Kwon \emph{et~al.}, ``Near-memory processing in action: Accelerating
  personalized recommendation with axdimm,'' \emph{IEEE Micro}, 2021.

\bibitem{nearmemory2}
S.~Khoram, Y.~Zha, J.~Zhang, and J.~Li, ``Challenges and opportunities: From
  near-memory computing to in-memory computing,'' in \emph{Proceedings of the
  2017 ACM on International Symposium on Physical Design}, 2017, pp. 43--46.

\bibitem{fault4}
T.~Kwon, M.~Imran, J.~M. You, and J.-S. Yang, ``Heterogeneous pcm array
  architecture for reliability, performance and lifetime enhancement,'' in
  \emph{2018 Design, Automation \& Test in Europe Conference \& Exhibition
  (DATE)}.\hskip 1em plus 0.5em minus 0.4em\relax IEEE, 2018, pp. 1610--1615.

\bibitem{kwon2019tensordimm}
Y.~Kwon, Y.~Lee, and M.~Rhu, ``Tensordimm: A practical near-memory processing
  architecture for embeddings and tensor operations in deep learning,'' in
  \emph{Proceedings of the 52nd Annual IEEE/ACM International Symposium on
  Microarchitecture}, 2019, pp. 740--753.

\bibitem{INM_database2}
T.~Lahiri, S.~Chavan, M.~Colgan, D.~Das, A.~Ganesh, M.~Gleeson, S.~Hase,
  A.~Holloway, J.~Kamp, T.-H. Lee \emph{et~al.}, ``Oracle database in-memory: A
  dual format in-memory database,'' in \emph{2015 IEEE 31st International
  Conference on Data Engineering}.\hskip 1em plus 0.5em minus 0.4em\relax IEEE,
  2015, pp. 1253--1258.

\bibitem{INM_database1}
T.~Lahiri, M.-A. Neimat, and S.~Folkman, ``Oracle timesten: An in-memory
  database for enterprise applications.'' \emph{IEEE Data Eng. Bull.}, vol.~36,
  no.~2, pp. 6--13, 2013.

\bibitem{lee2020bit}
K.~Lee, J.~Jeong, S.~Cheon, W.~Choi, and J.~Park, ``Bit parallel 6t sram
  in-memory computing with reconfigurable bit-precision,'' in \emph{2020 57th
  ACM/IEEE Design Automation Conference (DAC)}.\hskip 1em plus 0.5em minus
  0.4em\relax IEEE, 2020, pp. 1--6.

\bibitem{NVIDIA}
G.~Li, S.~K.~S. Hari, M.~Sullivan, T.~Tsai, K.~Pattabiraman, J.~Emer, and S.~W.
  Keckler, ``Understanding error propagation in deep learning neural network
  (dnn) accelerators and applications,'' in \emph{Proceedings of the
  International Conference for High Performance Computing, Networking, Storage
  and Analysis}, 2017, pp. 1--12.

\bibitem{li2016pinatubo}
S.~Li, C.~Xu, Q.~Zou, J.~Zhao, Y.~Lu, and Y.~Xie, ``Pinatubo: A
  processing-in-memory architecture for bulk bitwise operations in emerging
  non-volatile memories,'' in \emph{Proceedings of the 53rd Annual Design
  Automation Conference}, 2016, pp. 1--6.

\bibitem{mutlu2019processing}
O.~Mutlu, S.~Ghose, J.~G{\'o}mez-Luna, and R.~Ausavarungnirun, ``Processing
  data where it makes sense: Enabling in-memory computation,''
  \emph{Microprocessors and Microsystems}, vol.~67, pp. 28--41, 2019.

\bibitem{XED}
P.~J. Nair, V.~Sridharan, and M.~K. Qureshi, ``Xed: Exposing on-die error
  detection information for strong memory reliability,'' in \emph{2016 ACM/IEEE
  43rd Annual International Symposium on Computer Architecture (ISCA)}.\hskip
  1em plus 0.5em minus 0.4em\relax IEEE, 2016, pp. 341--353.

\bibitem{fault3}
J.~Park, M.~Jo, E.~M. Bourim, J.~Yoon, D.-J. Seong, J.~Lee, W.~Lee, and
  H.~Hwang, ``Investigation of state stability of low-resistance state in
  resistive memory,'' \emph{IEEE Electron Device Letters}, vol.~31, no.~5, pp.
  485--487, 2010.

\bibitem{inmemory2}
K.~Roy, I.~Chakraborty, M.~Ali, A.~Ankit, and A.~Agrawal, ``In-memory computing
  in emerging memory technologies for machine learning: an overview,'' in
  \emph{2020 57th ACM/IEEE Design Automation Conference (DAC)}.\hskip 1em plus
  0.5em minus 0.4em\relax IEEE, 2020, pp. 1--6.

\bibitem{ambit}
V.~Seshadri, D.~Lee, T.~Mullins, H.~Hassan, A.~Boroumand, J.~Kim, M.~A. Kozuch,
  O.~Mutlu, P.~B. Gibbons, and T.~C. Mowry, ``Ambit: In-memory accelerator for
  bulk bitwise operations using commodity dram technology,'' in \emph{2017 50th
  Annual IEEE/ACM International Symposium on Microarchitecture (MICRO)}.\hskip
  1em plus 0.5em minus 0.4em\relax IEEE, 2017, pp. 273--287.

\bibitem{seshadri2019dram}
V.~Seshadri and O.~Mutlu, ``In-dram bulk bitwise execution engine,''
  \emph{arXiv preprint arXiv:1905.09822}, 2019.

\bibitem{ISSAC}
A.~Shafiee, A.~Nag, N.~Muralimanohar, R.~Balasubramonian, J.~P. Strachan,
  M.~Hu, R.~S. Williams, and V.~Srikumar, ``Isaac: A convolutional neural
  network accelerator with in-situ analog arithmetic in crossbars,'' \emph{ACM
  SIGARCH Computer Architecture News}, vol.~44, no.~3, pp. 14--26, 2016.

\bibitem{sills2015high}
S.~Sills, A.~Calderoni, N.~Ramaswamy, S.~Yasuda, and K.~Aratani, ``High-density
  reram for storage class memory,'' in \emph{2015 15th Non-Volatile Memory
  Technology Symposium (NVMTS)}.\hskip 1em plus 0.5em minus 0.4em\relax IEEE,
  2015, pp. 1--4.

\bibitem{nearmemory1}
G.~Singh, L.~Chelini, S.~Corda, A.~J. Awan, S.~Stuijk, R.~Jordans,
  H.~Corporaal, and A.-J. Boonstra, ``A review of near-memory computing
  architectures: Opportunities and challenges,'' in \emph{2018 21st Euromicro
  Conference on Digital System Design (DSD)}.\hskip 1em plus 0.5em minus
  0.4em\relax IEEE, 2018, pp. 608--617.

\bibitem{PipeLayer}
L.~Song, X.~Qian, H.~Li, and Y.~Chen, ``Pipelayer: A pipelined reram-based
  accelerator for deep learning,'' in \emph{2017 IEEE international symposium
  on high performance computer architecture (HPCA)}.\hskip 1em plus 0.5em minus
  0.4em\relax IEEE, 2017, pp. 541--552.

\bibitem{26262}
P.~Stirgwolt, ``Effective management of functional safety for iso 26262
  standard,'' in \emph{2013 Proceedings Annual Reliability and Maintainability
  Symposium (RAMS)}.\hskip 1em plus 0.5em minus 0.4em\relax IEEE, 2013, pp.
  1--6.

\bibitem{inmemory1}
N.~Verma, H.~Jia, H.~Valavi, Y.~Tang, M.~Ozatay, L.-Y. Chen, B.~Zhang, and
  P.~Deaville, ``In-memory computing: Advances and prospects,'' \emph{IEEE
  Solid-State Circuits Magazine}, vol.~11, no.~3, pp. 43--55, 2019.

\bibitem{cong_xbar}
C.~Xu, D.~Niu, N.~Muralimanohar, R.~Balasubramonian, T.~Zhang, S.~Yu, and
  Y.~Xie, ``Overcoming the challenges of crossbar resistive memory
  architectures,'' in \emph{2015 IEEE 21st international symposium on high
  performance computer architecture (HPCA)}.\hskip 1em plus 0.5em minus
  0.4em\relax IEEE, 2015, pp. 476--488.

\bibitem{cong_impact}
C.~Xu, D.~Niu, Y.~Zheng, S.~Yu, and Y.~Xie, ``Impact of cell failure on
  reliable cross-point resistive memory design,'' \emph{ACM Transactions on
  Design Automation of Electronic Systems (TODAES)}, vol.~20, no.~4, pp. 1--21,
  2015.

\bibitem{yin2020xnor}
S.~Yin, Z.~Jiang, J.-S. Seo, and M.~Seok, ``Xnor-sram: In-memory computing sram
  macro for binary/ternary deep neural networks,'' \emph{IEEE Journal of
  Solid-State Circuits}, vol.~55, no.~6, pp. 1733--1743, 2020.

\bibitem{zabihi2018memory}
M.~Zabihi, Z.~I. Chowdhury, Z.~Zhao, U.~R. Karpuzcu, J.-P. Wang, and S.~S.
  Sapatnekar, ``In-memory processing on the spintronic cram: From hardware
  design to application mapping,'' \emph{IEEE Transactions on Computers},
  vol.~68, no.~8, pp. 1159--1173, 2018.

\bibitem{zhang2018exploring}
D.~Zhang, V.~Sridharan, and X.~Jian, ``Exploring and optimizing
  chipkill-correct for persistent memory based on high-density nvrams,'' in
  \emph{2018 51st Annual IEEE/ACM International Symposium on Microarchitecture
  (MICRO)}.\hskip 1em plus 0.5em minus 0.4em\relax IEEE, 2018, pp. 710--723.

\bibitem{zhang2019quantifying}
Z.~Zhang, L.~Huang, R.~Huang, W.~Xu, and D.~S. Katz, ``Quantifying the impact
  of memory errors in deep learning,'' in \emph{2019 IEEE International
  Conference on Cluster Computing (CLUSTER)}.\hskip 1em plus 0.5em minus
  0.4em\relax IEEE, 2019, pp. 1--12.

\end{thebibliography}


\end{document}